\documentclass[sigplan, nonacm]{acmart}

\pdfoutput=1

\usepackage[lined, noend, linesnumbered]{algorithm2e}
\SetKwBlock{Init}{}{}

\SetCommentSty{mycommfont}
\usepackage{tikz}
\usepackage{xcolor}
\usepackage{caption}
\usepackage{subcaption}
\usepackage{cleveref}
\newcommand{\name}{Chameleon}

\title{Towards Reconfigurable Linearizable Reads}

\author{Myles Thiessen}
\email{mthiessen@cs.toronto.edu}
\affiliation{
  \institution{University of Toronto}
  \city{Toronto}
  \country{Canada}
}

\author{Aleksey Panas}
\email{aleksey.panas@mail.utoronto.ca}
\affiliation{
  \institution{University of Toronto}
  \city{Toronto}
  \country{Canada}
}

\author{Guy Khazma}
\email{guykhazma@cs.toronto.edu}
\affiliation{
  \institution{University of Toronto}
  \city{Toronto}
  \country{Canada}
}

\author{Eyal de Lara}
\email{delara@cs.toronto.edu}
\affiliation{
  \institution{University of Toronto}
  \city{Toronto}
  \country{Canada}
}

\begin{document}

\begin{abstract}
    Linearizable datastores are desirable because they provide users with the illusion that the datastore is run on a single machine that performs client operations one at a time.
To reduce the performance cost of providing this illusion, many specialized algorithms for linearizable reads have been proposed which significantly improve read performance compared to write performance.
The main difference between these specialized algorithms is their performance under different workloads.
Unfortunately, since a datastore's workload is often unknown or changes over time and system designers must decide on a single read algorithm to implement ahead of time, a datastore's performance is often suboptimal as it cannot adapt to workload changes.

In this paper, we lay the groundwork for addressing this problem by proposing Chameleon, an algorithm for linearizable reads that provides a principled approach for datastores to switch between existing read algorithms at runtime.
The key observation that enables this generalization is that all existing algorithms are specific read-write quorum systems.
Chameleon constructs a generic read-write quorum system, by using \textit{tokens} that are \textit{included} to complete write and read operations.
This \textit{token quorum system} enables Chameleon to mimic existing read algorithms and switch between them by transferring these tokens between processes.
\end{abstract}

\maketitle

\section{Introduction}

A datastore is \textit{linearizable} if it provides users with the illusion that it is run on a single machine that processes client operations one at a time~\cite{herlihy1990linearizability}.
This property is desirable, as application developers do not need to reason about side effects that can arise from concurrent datastore operations~\cite{gunawi2014bugs, leesatapornwongsa2016taxdc}.

The de facto standard for providing linearizability is by use of a state machine replication algorithm~\cite{schneider1990implementing}.
These algorithms provide clients with procedures for both writing and reading the application's state.
While both write and read operations can be processed with the same procedure, the non-mutating property of read operations allows them to be processed concurrently without coordination between them.
This observation has led to the development of many specialized read algorithms which significantly improve read performance compared to write performance~\cite{charapko2019linearizable, chandra2007paxos, guarnieri2023linearizable, baker2011megastore, moraru2014paxos, katsarakis2020hermes, chandra2016algorithm, bi2022parameterized}.
These algorithms are of particular practical importance as many real-world workloads are read dominant~\cite{atikoglu2012workload, cheng2022taobench}.
For example, Google reported that in their ads workload, read operations outnumber write operations by three orders of magnitude~\cite{corbett2013spanner}.
This prevalence makes read operation performance the primary determinant of end-user latency and overall system throughput.

Existing read algorithms can be classified into one of four categories: leader~\cite{chandra2007paxos}, majority quorum~\cite{charapko2019linearizable, yi2023gleaning}, flexible quorum~\cite{guarnieri2023linearizable, whittaker2021read}, and local~\cite{baker2011megastore, moraru2014paxos, katsarakis2020hermes, chandra2016algorithm, bi2022parameterized}.
The main difference between algorithms in these categories is their performance in different workloads.
For example, in geo-distributed deployments, if most read operations are performed near a distinguished process known as the \textit{leader}, average read latency will be significantly lower using leader reads compared to majority quorum reads.
However, majority quorum reads can achieve higher peak throughput comparatively as the leader is not a bottleneck~\cite{arora2017leader, charapko2019linearizable}.
Moreover, local reads provide the lowest read latency compared to algorithms in all other categories at the cost of increased write latency~\cite{moraru2014paxos}.
Finally, flexible quorum reads increase read latency but reduce write latency in the presence of failures or network partitions.

These differences pose a serious challenge during the design of a linearizable datastore when the expected workload is unknown ahead of time.
Even in the unlikely scenario where it is known, designers can only implement a single read algorithm, and as such, performance will degrade if the workload changes.
Consequently, to optimize performance when the workload is unknown ahead of time and to adapt to workload changes, a datastore needs to be able to switch between existing read algorithms at run-time.

\def\heightgap{1}

\def\widthgap{.5}

\begin{figure*}
    \begin{subfigure}[b]{.24\textwidth}
        \centering
        \begin{tikzpicture}
            \tikzstyle{every node}=[font=\small]
            \begin{scope}
                \node[circle, draw, fill=white] (circle1) at (0, 2*\heightgap) {};
        
                \node[circle, draw, fill=white] (circle2) at (2*\widthgap, 1*\heightgap) {};
        
                \node[circle, draw, fill=white] (circle3) at (-2*\widthgap, 1*\heightgap) {};
        
                \node[circle, draw, fill=white] (circle4) at (-1*\widthgap, 0) {};
        
                \node[circle, draw, fill=white] (circle5) at (1*\widthgap, 0) {};
        
                \draw plot [smooth cycle,tension=1] coordinates {(.25, 2*\heightgap + .25) (-2*\widthgap - .25, 1*\heightgap) (-1*\widthgap + .25, -.25)};
        
                \draw [dashed] plot [smooth cycle,tension=1] coordinates {(0, 2*\heightgap + .25) (.25, 2*\heightgap) (0, 2*\heightgap - .25) (-.25, 2*\heightgap)};
            \end{scope}
        \end{tikzpicture}
        \caption{Leader reads.}
        \label{fig:leader_reads}
    \end{subfigure}
    \begin{subfigure}[b]{.24\textwidth}
        \centering
        \begin{tikzpicture}
            \tikzstyle{every node}=[font=\small]
            \begin{scope}
                \node[circle, draw, fill=white] (circle1) at (0, 2*\heightgap) {};
        
                \node[circle, draw, fill=white] (circle2) at (2*\widthgap, 1*\heightgap) {};
        
                \node[circle, draw, fill=white] (circle3) at (-2*\widthgap, 1*\heightgap) {};
        
                \node[circle, draw, fill=white] (circle4) at (-1*\widthgap, 0) {};
        
                \node[circle, draw, fill=white] (circle5) at (1*\widthgap, 0) {};
        
                \draw plot [smooth cycle,tension=1] coordinates {(.25, 2*\heightgap + .25) (-2*\widthgap - .25, 1*\heightgap) (-1*\widthgap + .25, -.25)};
        
                \draw [dashed] plot [smooth cycle,tension=1] coordinates {(-.25, 2*\heightgap + .25) (2*\widthgap + .25, 1*\heightgap) (1*\widthgap - .25, -.25)};
            \end{scope}
        \end{tikzpicture}
        \caption{Majority quorum reads.}
        \label{fig:quorum_reads}
    \end{subfigure}
    \begin{subfigure}[b]{.24\textwidth}
        \centering
        \begin{tikzpicture}
            \tikzstyle{every node}=[font=\small]
            \begin{scope}
                \node[circle, draw, fill=white] (circle1) at (0, 2*\heightgap) {};
        
                \node[circle, draw, fill=white] (circle2) at (2*\widthgap, 1*\heightgap) {};
        
                \node[circle, draw, fill=white] (circle3) at (-2*\widthgap, 1*\heightgap) {};
        
                \node[circle, draw, fill=white] (circle4) at (-1*\widthgap, 0) {};
        
                \node[circle, draw, fill=white] (circle5) at (1*\widthgap, 0) {};
        
                \draw plot [smooth cycle,tension=1] coordinates {(.25, 2*\heightgap + .25) (-2*\widthgap - .25, 1*\heightgap) (-1*\widthgap + .25, -.25)};
        
                \draw [dashed] plot [smooth cycle,tension=1] coordinates {(-1*\widthgap, .35) (1*\widthgap, .35) (1*\widthgap + .35, 0) (1*\widthgap, -.35) (-1*\widthgap, -.35) (-1*\widthgap -.35, 0)};
            \end{scope}
        \end{tikzpicture}
        \caption{Flexible quorum reads.}
        \label{fig:flexible_quorum_reads}
    \end{subfigure}
    \begin{subfigure}[b]{.24\textwidth}
        \centering
        \begin{tikzpicture}
            \tikzstyle{every node}=[font=\small]
            \begin{scope}
                \node[circle, draw, fill=white] (circle1) at (0, 2*\heightgap) {};
        
                \node[circle, draw, fill=white] (circle2) at (2*\widthgap, 1*\heightgap) {};
        
                \node[circle, draw, fill=white] (circle3) at (-2*\widthgap, 1*\heightgap) {};
        
                \node[circle, draw, fill=white] (circle4) at (-1*\widthgap, 0) {};
        
                \node[circle, draw, fill=white] (circle5) at (1*\widthgap, 0) {};

                \draw plot [smooth cycle,tension=.5] coordinates {(0, 2*\heightgap + .35) (-2*\widthgap - .35, 1*\heightgap) (-1*\widthgap - .25, -.25) (1*\widthgap + .25, -.25) (2*\widthgap + .35, 1*\heightgap)};

                \draw [dashed] plot [smooth cycle,tension=1] coordinates {(0, 2*\heightgap + .25) (.25, 2*\heightgap) (0, 2*\heightgap - .25) (-.25, 2*\heightgap)};

                \draw [dashed] plot [smooth cycle,tension=1] coordinates {(2*\widthgap, 1*\heightgap + .25) (2*\widthgap + .25, 1*\heightgap) (2*\widthgap, 1*\heightgap - .25) (2*\widthgap -.25, 1*\heightgap)};

                \draw [dashed] plot [smooth cycle,tension=1] coordinates {(-2*\widthgap, 1*\heightgap + .25) (-2*\widthgap + .25, 1*\heightgap) (-2*\widthgap, 1*\heightgap - .25) (-2*\widthgap -.25, 1*\heightgap)};
                
                \draw [dashed] plot [smooth cycle,tension=1] coordinates {(1*\widthgap, .25) (1*\widthgap + .25, 0) (1*\widthgap, -.25) (1*\widthgap -.25, 0)};

                \draw [dashed] plot [smooth cycle,tension=1] coordinates {(-1*\widthgap, .25) (-1*\widthgap + .25, 0) (-1*\widthgap, -.25) (-1*\widthgap -.25, 0)};
            \end{scope}
        \end{tikzpicture}
        \caption{Local reads.}
        \label{fig:local_reads}
    \end{subfigure}
    \caption{Example read (dashed) and write (solid) quorums of existing algorithms for linearizable reads.}
    \label{fig:quorums_of_existing_algorithms}
\end{figure*}
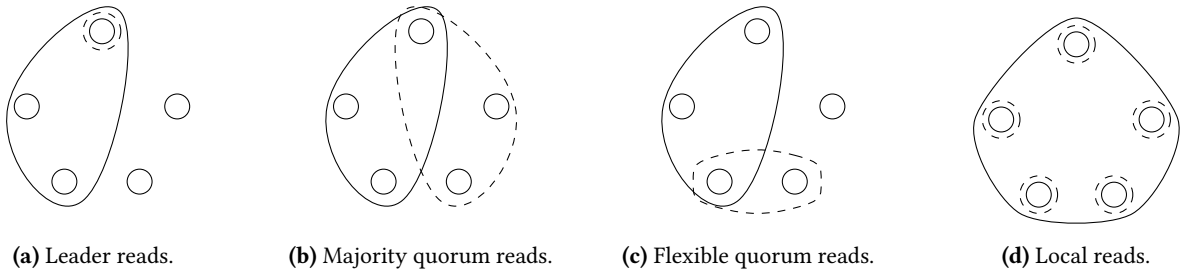

In this paper, we lay the groundwork for switching between different read algorithms at run-time by presenting \textit{\name{}}, a generalized and reconfigurable algorithm for linearizable reads.
The key observation that enables this generalization is that \textit{all existing algorithms are specific read-write quorum systems} --- their correctness is solely based on read and write operations contacting overlapping sets of processes.
To guarantee read-write quorum intersection, each process is provided with a set of \textit{tokens} that are \textit{included} to complete both read and write operations.
This \textit{token quorum system} enables \name{} to mimic existing read algorithms and switch between them by strategically transferring these tokens between processes.

In the remainder of this paper, we describe \name{} in detail.
\Cref{sec:background} argues that all existing algorithms are specific read-write quorum systems.
\Cref{sec:chameleon} describes the token quorum system and shows how it can mimic all existing algorithms.
\Cref{sec:discussion} discusses reconfiguring the token quorum system and tolerating failures.
\Cref{sec:related_work} discusses related work, and \Cref{sec:future_work} presents concluding remarks.

\section{Preliminaries}
\label{sec:background}

In this section, we discuss the system model, state machine replication, existing algorithms for linearizable reads, and their relationship to read-write quorum systems.

\subsection{Model}

We assume an asynchronous system consisting of $n$ processes of which at most $f$ may crash where $f < \frac{n}{2}$.
In this system, messages passed between processes may be arbitrarily delayed, reordered, or lost entirely.
In addition to this model, we require processes to be able to grant \textit{leases} in order to tolerate failures without sacrificing liveliness~\cite{gray1989leases}.
Specifically, we require that processes can grant leases \textit{correctly}, that is, the lease granter's perception of the lease's expiration is guaranteed to occur after the holder's perception of the lease's expiration in real-time.
This can be done by either assuming processes have access to synchronized clocks with bounded clock skew~\cite{gray1989leases} or by assuming processes are equipped with hardware clocks with bounded drift~\cite{liskov1991practical}.

\subsection{State Machine Replication}
\label{sec:smr}

The de facto standard for providing linearizable, fault-tolerant replication of client operations across a set of processes is by use of a state machine replication (SMR) algorithm.
Linearizability is a consistency model that ensures client operations are executed in some total order that respects their real-time ordering, i.e., if operation $o_1$ completes before operation $o_2$ begins, then $o_1$ must be ordered before $o_2$~\cite{herlihy1990linearizability}.

SMR algorithms provide fault-tolerance by maintaining a copy of the application's state at each process known as their \textit{replica}.
Replicas are assumed to atomically apply client operations and be deterministic, as in, if two replicas apply the same sequence of operations, they reach the same state~\cite{schneider1990implementing}.

Clients perform operations by submitting them to processes within the system.
Once a process receives such a request, it performs either the write or read procedure of the SMR algorithm depending on the submitted operation, and returns the result to the client.
Intuitively, the result for a write operation signifies the operation has taken effect and is durable whereas the result of a read operation is the latest version of the application's state.

To provide the real-time ordering guarantee of linearizability, SMR algorithms often rely on a distinguished process known as the \textit{leader} to sequence write operations.
The leader does so in two phases: prepare and commit.
The prepare phase consists of the leader proposing that a write operation $w$ be assigned to the $i$-th location in the \textit{log}.
The prepare phase completes once a majority of processes accept this proposal ensuring it is \textit{decided}.
After this, $w$ can be committed, that is, a process can apply $w$ to their replica after they apply all write operations assigned up to index $i$.

Read operations are processed by assigning them to the index number of either the latest committed write operation or some concurrent write operation.
Once a read operation has been assigned to some index number $i$, it is completed by executing it against some replica that has applied all write operations with index numbers up to and including $i$.

\def\heightgap{1}

\def\widthgap{.5}

\begin{figure*}
    \begin{subfigure}[b]{.24\textwidth}
        \centering
        \begin{tikzpicture}
            \tikzstyle{every node}=[font=\small]
            \begin{scope}
                \node[rectangle, draw, fill=white, label={above:5}] (rectangle1) at (0, 2*\heightgap) {A};
        
                \node[rectangle, draw, fill=white, label={right:0}] (rectangle2) at (2*\widthgap, 1*\heightgap) {C};
        
                \node[rectangle, draw, fill=white, label={left:0}] (rectangle3) at (-2*\widthgap, 1*\heightgap) {B};
        
                \node[rectangle, draw, fill=white, label={below:0}] (rectangle4) at (-1*\widthgap, 0) {D};
        
                \node[rectangle, draw, fill=white, label={below:0}] (rectangle5) at (1*\widthgap, 0) {E};
        
                \draw[->]
                    (rectangle2) edge (rectangle1)
                    (rectangle3) edge (rectangle1)
                    (rectangle4) edge (rectangle1)
                    (rectangle5) edge (rectangle1)
                ;
            \end{scope}
        \end{tikzpicture}
        \caption{Leader reads.}
        \label{fig:chameleon_leader_reads}
    \end{subfigure}
    \begin{subfigure}[b]{.24\textwidth}
        \centering
        \begin{tikzpicture}
            \tikzstyle{every node}=[font=\small]
            \begin{scope}
                \node[rectangle, draw, fill=white, label={above:1}] (rectangle1) at (0, 2*\heightgap) {A};
        
                \node[rectangle, draw, fill=white, label={right:1}] (rectangle2) at (2*\widthgap, 1*\heightgap) {C};
        
                \node[rectangle, draw, fill=white, label={left:1}] (rectangle3) at (-2*\widthgap, 1*\heightgap) {B};
        
                \node[rectangle, draw, fill=white, label={below:1}] (rectangle4) at (-1*\widthgap, 0) {D};
        
                \node[rectangle, draw, fill=white, label={below:1}] (rectangle5) at (1*\widthgap, 0) {E};
            \end{scope}
        \end{tikzpicture}
        \caption{Majority quorum reads.}
        \label{fig:chameleon_quorum_reads}
    \end{subfigure}
    \begin{subfigure}[b]{.24\textwidth}
        \centering
        \begin{tikzpicture}
            \tikzstyle{every node}=[font=\small]
            \begin{scope}
                \node[rectangle, draw, fill=white, label={above:1}] (rectangle1) at (0, 2*\heightgap) {A};
        
                \node[rectangle, draw, fill=white, label={right:1}] (rectangle2) at (2*\widthgap, 1*\heightgap) {C};
        
                \node[rectangle, draw, fill=white, label={left:0}] (rectangle3) at (-2*\widthgap, 1*\heightgap) {B};
        
                \node[rectangle, draw, fill=white, label={below:2}] (rectangle4) at (-1*\widthgap, 0) {D};
        
                \node[rectangle, draw, fill=white, label={below:1}] (rectangle5) at (1*\widthgap, 0) {E};
        
                \draw[->]
                    (rectangle3) edge (rectangle4)
                ;
            \end{scope}
        \end{tikzpicture}
        \caption{Flexible quorum reads.}
        \label{fig:chameleon_flexible_quorum_reads}
    \end{subfigure}
    \begin{subfigure}[b]{.24\textwidth}
        \centering
        \begin{tikzpicture}
            \tikzstyle{every node}=[font=\small]
            \begin{scope}
                \node[rectangle, draw, fill=white, label={above:5}] (rectangle1) at (0, 2*\heightgap) {A};
        
                \node[rectangle, draw, fill=white, label={right:5}] (rectangle2) at (2*\widthgap, 1*\heightgap) {C};
        
                \node[rectangle, draw, fill=white, label={left:5}] (rectangle3) at (-2*\widthgap, 1*\heightgap) {B};
        
                \node[rectangle, draw, fill=white, label={below:5}] (rectangle4) at (-1*\widthgap, 0) {D};
        
                \node[rectangle, draw, fill=white, label={below:5}] (rectangle5) at (1*\widthgap, 0) {E};
        
                \draw[->]
                    (rectangle1) edge (rectangle2)
                    (rectangle1) edge (rectangle3)
                    (rectangle1) edge (rectangle4)
                    (rectangle1) edge (rectangle5)
                    (rectangle2) edge (rectangle1)
                    (rectangle2) edge (rectangle3)
                    (rectangle2) edge (rectangle4)
                    (rectangle2) edge (rectangle5)
                    (rectangle3) edge (rectangle1)
                    (rectangle3) edge (rectangle2)
                    (rectangle3) edge (rectangle4)
                    (rectangle3) edge (rectangle5)
                    (rectangle4) edge (rectangle1)
                    (rectangle4) edge (rectangle2)
                    (rectangle4) edge (rectangle3)
                    (rectangle4) edge (rectangle5)
                    (rectangle5) edge (rectangle1)
                    (rectangle5) edge (rectangle2)
                    (rectangle5) edge (rectangle3)
                    (rectangle5) edge (rectangle4)
                ;
            \end{scope}
        \end{tikzpicture}
        \caption{Local reads.}
        \label{fig:chameleon_local_reads}
    \end{subfigure}
    \caption{Examples of \name{} mimicking existing algorithms for linearizable reads by transferring tokens between processes (represented by arrows). The number beside each process represents how many tokens it holds.}
\end{figure*}
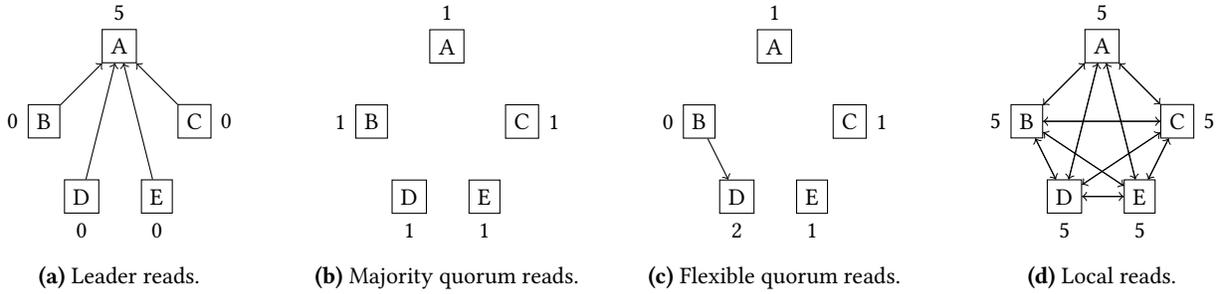

\subsection{Algorithms for Linearizable Reads}

We now describe the four categories of read algorithms mentioned previously.
In leader reads all read operations are forwarded to the network's leader for processing.
When the leader receives a read request it assigns it to the highest index that it has sent a commit request for.
This guarantees that read operations always observe the latest complete write operations so long as a new leader has not been elected in the meantime.
To ensure this doesn't happen, the leader is granted a \textit{leader lease}, guaranteeing at most one process believes they are the leader at any given time.
Unlike leader reads, in majority quorum reads, all processes can perform read operations.
They do so by assigning read operations to the maximum index a simple majority of processes have received a prepare request for.
Flexible quorum reads are similar to majority quorum reads except reads are assigned to the maximum index of an arbitrary read quorum of processes.
To enable this flexibility and guarantee the real-time ordering requirement of linearizability, the leader waits to receive acknowledgments from at least one process in every read quorum before committing a write.
Similarly to leader reads, to not reduce fault-tolerance, processes that are members of a read quorum are granted \textit{read leases}.
Finally, in local reads, all processes assign read operations to an index based on their local perception of the latest assigned write index.
Similar to flexible quorum reads, the leader in local reads must receive acknowledgments from all processes that hold a read lease before a write is committed.

\subsection{Read-Write Quorum Systems}

We now discuss the relationship between existing algorithms for linearizable reads and read-write quorum systems.
We do so with the aid of the illustration in \Cref{fig:quorums_of_existing_algorithms}.
A read-write quorum system is defined by a set of read quorums and write quorums such that every read and write quorum intersects.
In the context of specialized algorithms for linearizable reads, each read and write quorum is a subset of the $n$ processes in the system.
In leader reads, the only read quorum is the leader and any simple majority of processes including the leader is a write quorum.
Since at most one process is the leader at any time, every read and write quorum intersects.
In majority quorum reads, read and write quorums are any simple majority of processes.
Similarly, since any two simple majorities intersect so does every read and write quorum.
Flexible quorum reads define the quorum system explicitly.
Finally, in local reads, every process is a read quorum and there is a single write quorum containing all processes.
Consequently, \textit{all existing algorithms for linearizable reads reduce to specific read-write quorum systems.}

\section{\name{}}
\label{sec:chameleon}

We now explain how \name{} constructs a read-write quorum system using \textit{tokens}, followed by showing how this enables \name{} to mimic all existing specialized read algorithms.
We then describe how \name{} performs write and read operations assuming that messages aren't lost, all processes are non-faulty, and the quorum system and leader are fixed followed by sketching why \name{} is correct.
We discuss relaxing these assumptions in \Cref{sec:discussion}.

\subsection{The Token Quorum System}
\label{sec:stationary_tokens}

Each token is \textit{owned} by one process and \textit{held} by at most one process at any time.
Note, a token's owner \textit{never} changes but its holder can.
We say that a token is \textit{transferred} from one process to another.
A token is defined as a tuple $(l, r)$ where $l$ is the token's owner and $r$ is some unique integer used to differentiate between tokens owned by the same process.
Using these tokens a read quorum is then defined by a set of processes that hold at least \textit{one} token owned by some simple majority of processes.
Similarly, a write quorum is defined by a set of processes which is at least a simple majority of processes that collectively hold \textit{every} token owned by some (potentially different) simple majority of processes.
Finally, we say that an operation \textit{included} a token if it completes by contacting a quorum that held it. 

\subsection{Mimicking Existing Algorithms}
\label{sec:mimicking_existing_algorithms}

We now show how strategically assigning which processes hold which tokens enables \name{} to mimic each of the existing read algorithms discussed in \Cref{sec:background}.

To mimic leader reads, each process owns a single token which is held by the leader.
This is illustrated in \Cref{fig:chameleon_leader_reads} where the arrows represent the transferring of tokens owned by processes B, C, D, and E to A (the leader), and the number beside each process represents how many tokens it holds.
In this scenario, the leader itself is a read quorum as it holds at least one token from a simple majority of processes.
Furthermore, a write quorum is any simple majority of processes that includes the leader --- the only process that holds every token from a majority of processes.

To mimic majority quorum reads each process owns a single token which is not transferred.
Consequently, any simple majority of processes is both a read and write quorum --- the same quorum system produced by majority quorum reads.
We illustrate an example of this in \Cref{fig:chameleon_quorum_reads}.

\name{} mimics flexible quorum reads through a combination of the previous two approaches, as only one process needs to hold more than one token to achieve a read quorum smaller than a simple majority.
We illustrate such an example in \Cref{fig:chameleon_flexible_quorum_reads} where process D holds two tokens.
This enables D and processes A, C, or E to perform read operations.
Furthermore, in this example, since a write must include every token owned by a simple majority of processes, it is bound to intersect D or A, C, and E.
In either case, this guarantees read-write quorum intersection.

Finally, \name{} mimics local reads by each process owning $n$ tokens and transferring one token to all other processes.
Consequently, each process will hold onto one token owned by every process in the system as we illustrate in \Cref{fig:chameleon_local_reads}.
This results in each process being a read quorum and the only valid write quorum being all processes. 
Consequently, the token quorum system can mimic \textit{every} existing specialized algorithm for linearizable reads.

\subsection{Performing Client Operations}
\label{sec:performing_client_operations}

We now discuss how \name{} performs operations using the token quorum system.

\begin{algorithm}[t]
\DontPrintSemicolon

\Init(\textbf{procedure} \text{write($o$):}){
    $cntr \coloneqq cntr + 1$\tcc*[f]{Local op counter initially 0}

    \tcc{$cntr$ on lines 3 and 4 is the same as line 2}
    \textbf{send} $\langle WRITE, o, cntr \rangle$ to $\ell$
    
    \textbf{wait to receive} $\langle WRITE\_ACK, cntr \rangle$ from $\ell$
}

\BlankLine

\Init(\textbf{upon receiving} $\langle WRITE, o, c \rangle$ from $p$:){
    $i \coloneq i + 1$\tcc*[f]{Latest log index initially 0}

    \textbf{send} $\langle PREPARE, i \rangle$ to all

    \tcc{Processes that have acknowledged, the tokens they have returned, and the current include.}
    $A \coloneqq TR \coloneqq TI \coloneqq \emptyset$

    \Init(\textbf{upon receiving} $\langle P\_ACK, i, T' \rangle$ from $q$:){
        $A \coloneqq A \cup \{q\}$ and $TR \coloneq TR \cup T'$

        \ForEach{process $r$ in the system}{
            \If{$TR$ contains $k$ tokens owned by $r$}{
                $TI \coloneqq TI \cup \{r\}$
            }
        }
    }

    \textbf{wait for} $|A| \geq \lceil \frac{n + 1}{2} \rceil$ and $|TI| \geq \lceil \frac{n + 1}{2} \rceil$

    \textbf{send} $\langle COMMIT, i, o \rangle$ to all

    \textbf{send} $\langle WRITE\_ACK, c \rangle$ to $p$
}

\BlankLine

\Init(\textbf{upon receiving} $\langle PREPARE, i \rangle$ from $\ell$:){
    \tcc{Maximum prepare index received initially 0}
    $MaxP \coloneqq \max(MaxP, i)$

    \tcc{$T$ is the set of held tokens.}
    \textbf{send} $\langle P\_ACK, i, T \rangle$ to $\ell$
}

\BlankLine

\Init(\textbf{upon receiving} $\langle COMMIT, i, o \rangle$ from $\ell$:){
    \tcc{Applies the write $o$ against the local replica once all writes up to $i - 1$ have been applied}
    \textit{Write($i$, $o$)}
}

\caption{Write pseudocode for \name{} assuming no message loss, all processes are non-faulty, a fixed leader $\ell$, each process holds a fixed set of tokens $T$, and each process owns $k$ tokens.}
\label{alg:write_procedure}
\end{algorithm}

\noindent\textbf{Writes (\Cref{alg:write_procedure}).} When a process wishes to perform a write operation, it first forwards it to the leader.
Once the leader receives this request, it assigns the write operation to the next index in the log and sends a prepare request to all processes including the write operation and its assigned index number.
Upon receipt of this request, a process records this index number if it is the highest they have received a prepare request for.
It then sends an acknowledgment to the leader including a description of the current set of tokens they hold.
In the system illustrated in \Cref{fig:chameleon_flexible_quorum_reads}, the set of tokens returned per process differs.
For example, processes A, C, and E will inform the leader they hold their own token, process B will return saying they do not hold any tokens, and process D will return saying they hold both their own and B's token.
Once the leader has sent these prepare requests, it waits to receive acknowledgments from at least a simple majority of processes that collectively hold every token owned by some (potentially different) simple majority of processes.
In the system illustrated in \Cref{fig:chameleon_flexible_quorum_reads}, valid sets of acknowledging processes include (A, C, E), (A, D, E), and (C, D, E).
Following this, the leader sends commit requests to all processes for this write and returns to the client signifying that the write is completed and durable.

\begin{algorithm}[t]
\DontPrintSemicolon

\Init(\textbf{procedure} \text{read($o$):}){
    $cntr \coloneqq cntr + 1$ and $index \coloneqq 0$
    
    $R \coloneqq \text{closest\_read\_quorum()}$

    \If{$R$ is only the current process}{
        $index \coloneqq MaxP$
    }
    \Else{
        \tcc{$cntr$ on lines 7 and 9 is the same as line 2}
        \textbf{send} $\langle READ, cntr \rangle$ to all processes in $R$
    
        $TI \coloneqq \emptyset$
    
        \Init(\textbf{upon receiving} $\langle R\_ACK, cntr, T', MaxP' \rangle$:){
            $index \coloneqq \max(index, MaxP')$
    
            \ForEach{process $p$ in the system}{
                \lIf{$(p, *) \in T'$}{
                    $TI \coloneqq TI \cup \{p\}$
                }
            }
        }
    
        \textbf{wait for} $|TI| \geq \lceil \frac{n + 1}{2} \rceil$
    }
    \tcc{Executes the read request $o$ against the local replica once all writes up to and including $index$ have been applied and return the result}
    \textbf{return} \textit{Read($index$, $o$)}.
}

\BlankLine

\Init(\textbf{upon receiving} $\langle READ, c \rangle$ from $p$:){
    \textbf{send} $\langle R\_ACK, c, T, MaxP \rangle$ to $p$
}

\caption{Read pseudocode for \name{} under the same assumptions as \Cref{alg:write_procedure}.}
\label{alg:read_procedure}
\end{algorithm}

\noindent\textbf{Reads (\Cref{alg:read_procedure}).} Similarly to write operations, clients submit read operations to any process.
Once a process $p$ receives this request, it first computes the closest read quorum to it ($R$) based on the tokens each process holds.
If this quorum happens to be $p$ itself, then $p$ assigns the read to the highest index it has received a prepare request for.
Otherwise, $p$ sends a \textit{read} request to all processes in $R$.
For example, if process $p$ is process A in \Cref{fig:chameleon_flexible_quorum_reads} possible $R$'s include (A, C, E), (A, D), (C, D), and (D, E).
Once a process receives a read request it returns the highest index it has received a prepare request for and the set of tokens it currently holds.
Once $p$ has sent these read requests, it waits to receive acknowledgments from a set of processes that collectively hold at least one token owned by some simple majority of processes.
Following this, $p$ assigns the read operation to the maximum index returned from this set of processes.
In either case,  $p$ then completes the read operation by (1) waiting for its replica to commit all write operations up to and including the assigned index, (2) executing the read operation against its local replica, and (3) returning the result to the client.

Sending read requests to the closest read quorum is similar to the \textit{thrifty} optimization of Paxos~\cite{lamport2004cheap} which only sends prepare messages to a majority of processes.
This reduces the number of messages sent by the leader but increases tail latency when processes don't respond.
To avoid this trade-off, $R$ can be replaced with all processes in the system.

\subsection{Correctness Sketch}
\label{sec:correctness}

\name{} is similar to MultiPaxos~\cite{lamport2001paxos} and Raft~\cite{moraru2014paxos} in the sense that the leader processes writes in two phases and assigns writes to log indexes in increasing order.
Consequently, the argument for \name{}'s correctness is fundamentally similar to these algorithms and as such we focus on its sole difference --- how reads are ordered with respect to writes.
Specifically, \name{} must guarantee that if a write operation $w$ assigned to index $i$ completes before some read operation $r$ begins then $r$ is assigned to an index at least $i$.
This follows from our assumption that the token quorum system is fixed and the fact that every read and write quorum has at least one token in their intersection.
These together imply that some process $p$ was in the intersection of $w$'s write quorum and $r$'s read quorum. 
Consequently, since $w$ completed before $r$ began, $p$ processed the prepare request for $w$ before the read request for $r$, and therefore $p$ would have returned a maximum prepare index of at least $i$ in response to the read request for $r$.
This in turn implies that $r$ is assigned to at least $i$ as required.

\section{Discussion}
\label{sec:discussion}

We now discuss mechanisms for reconfiguring the token quorum system and tolerating message loss and failures.

\subsection{Reconfiguring the Token Quorum System}

\name{}'s novelty lies in its ability to switch between existing algorithms for linearizable reads at runtime.
This is achieved by transferring tokens between processes --- reconfiguring the token quorum system in the process.
To maintain correctness \name{} must ensure that all read operations observe all previously completed write operations.
This can be achieved by ensuring that if a process holds some token $t$ it has a maximum prepare index at least as large as the maximum index number of any completed write operation that included $t$.
We now discuss a synchronous centralized approach to guarantee this.
For future work, we plan to explore how to minimize reconfiguration times by designing an asynchronous approach for reconfiguration.

In the centralized approach, the leader appends special entries to the log known as \textit{token configurations} that describe what processes should hold which tokens.
Processes then determine the current set of tokens they hold based on the latest configuration.
To not impact performance, each process knows the latest configuration locally.
To ensure this \textit{local perception} of the token configuration is consistent, processes mark their local perception as invalid once they receive a prepare request for a new token configuration.
At this point, processes do not know which tokens they hold and as such, their processing of prepare and read requests is stalled until they know.
This occurs after they receive a commit request for the new token configuration, which is only sent once the leader receives acknowledgments from all processes.

To guarantee that all read operations observe all previously completed write operations we need to make two additional changes.
First, when the leader processes a new token configuration, it (1) waits for all outstanding write operations to complete before proposing the token configuration and (2) stalls the processing of new write operations until the new token configuration has been acknowledged by all processes.
Second, when acknowledging read requests, processes return the index of the token configuration it used to compute the set of tokens it holds.
The process performing the read operation then keeps track of the highest index returned and counts tokens only from processes that return that index.
In the event this is insufficient to cover a read quorum, the process resends read requests until it covers a read quorum.
These changes ensure that each token holder has a maximum prepare index of at least as large as the maximum index number of any completed write operation.
This is because the new token configuration's index number is higher than all previously completed writes since the leader waited for all outstanding writes to be completed.
Furthermore, read operations are bound to observe all previously completed write operations since readers only include tokens from the latest token configuration.

\subsection{Tolerating Message Loss \& Failures}

Like in existing algorithms, message loss can be tolerated by supporting request re-transmission and detecting duplicate operations.
Consequently, \name{} can be adapted to do so using existing techniques such as the ones proposed by CHT~\cite{chandra2016algorithm}.
% We plan to formalize this adaption for future work.
Similarly, tolerating failures in \name{} is functionally equivalent to how algorithms for local reads tolerate failures.
This is because they both have to guarantee liveliness even when a read quorum is partitioned from the rest of the network.
To handle this case, algorithms for local reads grant processes \textit{read leases} which are time-based contracts that while valid allow processes to perform read operations against their local replica.
When these leaseholders crash or are partitioned from the rest of the network, the leader will wait for the expiration of their read lease before completing ongoing write operations.
This guarantees they are no longer performing reads locally and thus it is safe for the rest of the processes to commit said write.
We now discuss how this idea can be adapted to \name{} to tolerate failures.

To enable fault-tolerant local reading of the token configuration, each process is granted a read lease which is revoked when a process doesn't respond to token configuration changes promptly.
Similarly, to tolerate the failure of a token holder, holding a token is attached to a lease.
Specifically, if process $p$ holds a token $t$ owned by process $q$, $q$ will grant $p$ the right to hold $t$ for some lease duration.
This enables the leader to revoke all tokens held by some process $p$ by contacting all processes and requesting them to wait for their leases granted to $p$ to expire.
Once this occurs, the leader includes these revoked tokens until the write quorum condition is met.
Like in algorithms for local reads, this ensures that any read quorum that $p$ was a part of is no longer performing read operations after its leases are revoked.

In addition to this revocation mechanism, once a token has been revoked, the token needs to be associated with a valid prepare index to ensure read operations observe all completed write operations.
Like in our reconfiguration mechanism, this can be accomplished by relying on the leader to provide their latest assigned log index to the process which is revoking the token.
However, unlike the reconfiguration mechanisms, this requires that at most one process is the leader at any time.
This can be accomplished by granting the leader a leader lease.
To also handle the failure of the leader, we can adopt the enhanced leader election service proposed by CHT~\cite{chandra2016algorithm} that takes care of both granting the leader a leader lease and electing a leader.
For future work, we will formalize the mechanism required to tolerate failures.

\section{Related Work}
\label{sec:related_work}

\noindent\textbf{SMR algorithms.} The most well known SMR algorithms are MultiPaxos~\cite{lamport2001paxos} and Raft~\cite{ongaro2014search}.
Like \name{}, they are leader-based and process writes in two phases.
An alternative to these leader-based algorithms is leaderless algorithms such as EPaxos~\cite{moraru2013there, tollman2021epaxos} and Nezha~\cite{geng2022nezha}.
These algorithms allow all processes to perform write operations without contacting the leader at the expense of needing a \textit{super quorum} of acknowledgments instead of a simple majority.
We believe \name{} can be adapted to work with these algorithms.

\noindent\textbf{Quorum systems.} Flexible~\cite{howard2016flexible}, Wan~\cite{ailijiang2017multileader}, and Dynamic Paxos~\cite{nawab2018dpaxos} use non-majority quorum systems to ensure that leader election quorums always intersect with prepare quorums.
This guarantees that writes are consistent across leaders.
In Flexible and Wan Paxos, the quorum system is configurable but must be known a priori.
Dynamic Paxos allows for the quorum system to be reconfigured at runtime but is specific to leader election quorums.
Specifically, Dynamic Paxos enables the \textit{leader election zone} to move at runtime to tolerate failures or for access locality.
Due to the specific use case of this mechanism, it is not capable of mimicking existing algorithms for linearizable reads.
\name{}, on the other hand, can because of its use of the token quorum system.
EdgePQR~\cite{guarnieri2023linearizable} extends Wan Paxos with the ability to perform low-latency reads at edge data centers.
To do so, they introduce the idea of \textit{edge quorums} which enables reads to be processed by contacting any majority of processes within an edge datacenter.
To guarantee correctness, all write operations are also required to contact a majority of processes in each edge data center.
EdgePQR tolerates the failure of edge datacenters without stalling write operations by enabling the addition and removal of edge datacenters at runtime.
This is similar to \name{}'s ability to reconfigure the read-write quorum system but only applies to edge quorums and not the entire read-write quorum system.
Quoracle~\cite{whittaker2021read} quantifies the trade-offs between different read-write quorum systems but does not consider reconfiguration at runtime.

\noindent\textbf{Token virtualization.} An alternative interpretation of the token quorum system presented in this paper is that each token is a \textit{virtual} process which is mapped to a single \textit{physical} process through token holding.
Quorum intersection in \name{}'s case is then guaranteed by intersecting sufficiently many virtual processes for both reads and writes.
This idea of \textit{token virtualization} has parallels to works in the late 80s and early 90s done on fair resource scheduling~\cite{waldspurger1994lottery, kay1988fair}.
In these works, tokens were known as tickets or shares which were given to users to fairly schedule their programs on shared CPU resources.
\name{} differs from these works in its use of tokens as a means to guarantee read-write quorum intersection instead of fair scheduling.

\section{Conclusion}
\label{sec:future_work}

In this paper, we presented \name{}, an algorithm for linearizable reads that lays out a principled approach for datastores to switch between existing read algorithms at runtime.
To do so, \name{} constructs a read-write quorum system using tokens that are included to complete read and write operations.
This token quorum system enables \name{} to switch between existing read algorithms by transferring them between processes.
For future work, we plan to develop a complete description of \name{} along with an evaluation that demonstrates the benefits of switching between existing read algorithms at runtime.

\bibliographystyle{plain}
\bibliography{main}

\end{document}